\documentclass[twocolumn,showpacs,aps,floatfix]{revtex4}

\usepackage{amsmath}
\usepackage{graphicx}
\usepackage{dcolumn}
\usepackage{bm}

\begin{document}

\title{CI-RMBPT calculations of photoionization cross sections from quasi-continuum oscillator strengths }
\author{I. M. Savukov}
\affiliation{Los Alamos National Laboratory, Los Alamos, NM 87545,
USA}
\author{D. V. Filin}
\affiliation{New Mexico Consortium, Los Alamos, NM 87544}

\date{\today}

\begin{abstract}
Many applications are in need of accurate photoionization
cross-sections, especially in the case of complex atoms.
Configuration-interaction relativistic many-body theory (CI-RMBPT)
that has been successful in predicting atomic energies, matrix
elements between discrete states, and other properties is quite
promising, but it has not been applied to photo-ionization problems
owing to extra complications arising from continuum states. In this
paper a method that will allow the conversion of discrete CI-(R)MPBT
oscillator strengths (OS) to photo-ionization cross sections with
minimal modifications of the codes is introduced and CI-RMBPT cross
sections of Ne, Ar, Kr, Xe are calculated. A consistent agreement
with experiment is found. RMBPT corrections are particularly
significant for Ar, Kr, and Xe and improve agreement with experiment
compared to the particle-hole CI method. The demonstrated conversion
method can be applied to CI-RMBPT photo-ionization calculations for
a large number of multi-valence atoms and ions.
\end{abstract}
\pacs{32.80.Fb, 31.10.+z, 31.15.A-, 31.30.Jv}
 \maketitle

Many applications, such as plasma modeling and opacity calculations,
are in need of accurate photoionization cross-sections, especially
in the case of complex atoms. Fairly simple approaches, such as
random-phase approximation (RPA), have been applied to calculations
of photo-ionization cross-sections in alkali-metal \cite{JohnsonRPA}
and noble-gas atoms \cite{RPANoble}, but they are not adequate for
an arbitrary atom. A more general R-matrix approach has been used,
but owing to the complexity of implementation, it is not easy to
incorporate into existing precision atomic-structure methods to
systematically improve accuracy. Configuration-interaction
relativistic many-body theory (CI-RMBPT) formalism that has been
successful in predicting energies, matrix elements, and other
properties of bound states of atoms \cite{CIMBPTDzuba,CIMBPTNeon,
CIMBPTSafronova,CIMBPTAr} is quite promising, but it has not applied
to photo-ionization problems owing to extra complications arising
from continuum states. In this letter a method is introduced to
convert discrete CI-(R)MPBT oscillator strengths (OS) to
photo-ionization cross sections with minimal modifications of the
codes and CI-RMBPT cross sections are computed.

The proposed conversion method is the generalization of the previous
approach given in Ref.\cite{quasiPhoto} that consisted of placing an
atom in a cavity, calculating oscillator strenghts $f_{i}$ to
quasi-continuum states, and converting them to the differential
cross section $\sigma(\epsilon)$ using the relation:
\begin{equation}
\sigma(\epsilon)=4.03\times 10^{-18}\frac{df(\epsilon)}{d\epsilon}.
\end{equation}
The derivative was determined from discrete energies of
quasi-continuum states $\epsilon_i$, also available in the
calculations,
\begin{equation}
\frac{df(\epsilon)}{d\epsilon}\approx \frac{2
f_{i}}{\epsilon_{i+1}-\epsilon_{i-1}}.
\end{equation}

The derivative can be calculated quite accurately using Eq.(2) for
monovalent atoms, but large uncertainty appears  in multi-valence
atoms. This is the main difficulty for the conversion. Fortunately
with the specific statistical averaging over multiple configurations
to be described here fairly accurate cross sections still can be
obtained.

It is notable that a cavity is  also used in R-matrix methods
\cite{Rmatrix}; however, many important differences exist
\cite{Scatter} between the methods. In particular, the method
presented here does not require modifications of the CI-(R)MBPT
codes, while the R-matrix approach would lead to significant
changes, which would be quite very difficult to make in CI-MBPT
programs considering the complexity of the codes developed for
multi-valence atoms \cite{CIMBPTDzuba}. The advantage of the
R-matrix approach, on the other hand, is that the conversion
procedure described here is not required since continuum
wavefunctions are properly normalized. If the R-matrix approach were
properly incorporated into the CI-MBPT formalism, better accuracy
could be achieved. The current work can be considered as the first
step for the development of the R-matrix CI-MBPT method.

 We illustrate the accuracy of the method proposed here on the example of
CI-RMBPT calculations of cross sections of Ne, Ar, Kr, and Xe. These
atoms are chosen for the following reasons: 1) noble-gas atom
photoionization cross sections have been measured with high
precision; 2) the number of particle-hole configurations is
relatively small, facilitating the investigation of the effects of
various parameters on the accuracy; 3) an accurate particle-hole
CI+MBPT method has been previously developed for these atoms with
agreement demonstrated for energies, oscillator strengths, and
g-factors \cite{CIMBPTAr}; 4) these atoms are
 of considerable interest for applications.

In current calculations, we use the particle-hole (PH) CI-RMPBT
method described previously \cite{CIMBPTAr}. The CI-RMBPT terms are
evaluated using radial B-spline basic sets. To generate the basic
functions for calculations of photo-ionization cross-sections, the
following steps were implemented.  First, the Dirac-Hartree-Fock
(DHF) potential was generated for a closed-shell atom, such as Ne,
Ar, Kr, and Xe. Next, a B-spline subroutine was used to obtain basic
sets for the core and virtual states in the frozen DHF ground-state
potential. The spherical cavity ``bag" boundary condition
$P(R)=Q(R)$, where $P(r)$ and $Q(r)$ are the large and small
components of the radial Dirac wave function, was imposed to make
this basis discrete. The basis consisted of 40-100 radial functions
for each spin-orbit index, with the maximum orbital angular momentum
restricted to 5. The continuum wavefunctions were replaced with
quasicontinuum orbitals. Only minimal modifications were made of
existing CI-RMBPT codes and auxiliary subroutines, including the
B-spline subroutine. Namely, the maximum number of spline functions
was increased from 40 to 100, to improve numerical accuracy, which
was estimated from comparison of results when both the number of
splines and the cavity size were varied.

Particle-hole interactions in noble-gas atoms are strong and have to
be treated in all orders using the CI approach \cite{CIMBPTNeon}.
For this, the PH state functions
\begin{equation}
\begin{array}{c}
\Phi_{JM}(av)=\displaystyle\sum_{m_a m_v}(-1)^{j_v-m_v}(2J+1)^{1/2}
\left(
\begin{array}{c c c } j_v & J & j_a \\ -m_v & M & m_a
\end{array} \right)
\\
\times a_{vm_v}^\dag a_{am_a}
\end{array}
\end{equation}
are introduced for the construction of the CI matrix
$H[a'v'(J),av(J)]$. Here $J$ is the total coupled angular momentum,
$M$ is its projection, $j_v$ and $m_v$ are the angular momentum and
its projection of the excited state $v$, $j_a$ and $m_a$ are the
angular momentum and its projection of the hole state. The effective
Hamiltonian, which in the case of CI-RMBPT contains RMBPT
corrections, is evaluated between PH state functions and
diagonalized to obtain the expansion coefficients of the coupled
excited states of noble-gas atoms. It is known that the
particle-hole CI approach, restricted to single excitations, does
not provide very high accuracy, and it is necessary to take into
account particle-core and hole-core interactions. In principle the
CI basis can be extended to include double and triple excitations,
but the basis becomes large. Alternatively, following the CI-MBPT
method, MBPT corrections are incorporated to account for most
important interactions beyond the single-excitation CI. We include
second-order effects and some important higher-order effects by
modifying energies in the denominators. Such an approach has been
successful for predicting a large number of excited states and
transitions \cite{CIMBPTNeon,CIMBPTAr}.

The quasi-continuum PH spectrum generated in CI-(R)MBPT or CI
calculations is quite irregular and the cross section calculated
with Eqs. (1-2) has poor accuracy. Even averaging over a large
number of points does not solve the problem, leading to the loss of
resolution on the energy scale. Below we will demonstrate an
alternative method of calculations of the cross section
 based on the fitting of the sums of oscillator strengths and the differentiation of the fitted curve.

The rationale for this approach is the following. If the
differential cross section is integrated over energy, it becomes
related (compare Eq.(1)) to the sums of discrete quasi-continuum OSs
\begin{equation}
I(E_i)=\int_{E_0}^{E_i}{\sigma(E)dE}\approx F_i\equiv
4.03\sum_{j\leq i}{f_j}+C
\end{equation}
(Here, C is the constant accounting for the contribution of bound
discrete states, $E_0$ is the ionization energy, the photoionization
cross section is in Mb units.) Because now there is no division by
energy intervals, the accuracy of theoretical results not affected
by the irregularity of energy level is expected adequate as it is
illustrated for Ne (Fig. 1). (Note that a substantial difference
exists between two fairly reliable Ne experimental measurements
\cite{PrecNoble} and \cite{ChanEtAl2}. We chose the former, since
the precision claimed there is
 1-3\%, higher than in the latter, 5\%, estimated for the method in the follow up
 publication
 \cite{ChanEtAl3}.)
The method of comparison of the integrated cross sections and the
sums of quasi-continuum OSs can already be useful in its own right
for tests of theory and experimental results; however, it is also
possible to convert the sums to differential cross section. Now of
course energy intervals will reappear in the problem, and we have to
find a way to reduce ``noise." Differentiation of noisy data is an
ill-posed problem in general, but it is well-known that smoothing
can dramatically improve accuracy. By comparing cross sections
obtained by the differentiation of various polynomial fittings of
data subsets, we found that the line fitting method is sufficiently
accurate. Thus this method is chosen in our final calculations, with
the advantage of simple analytical solution. In this case the cross
section at the energy $E_l$ is:
\begin{equation}
\sigma(E_l)=\frac{(2N+1)\displaystyle\sum_{i=l-N}^{l+N}{E_i
F_i}-\displaystyle\sum_{i=l-N}^{l+N}{E_i}\sum_{i=l-N}^{l+N}{F_i}}{(2N+1)\displaystyle\sum_{i=l-N}^{l+N}{E_i^2}-(\sum_{i=l-N}^{l+N}{E_i})^2}
\end{equation}
where $E_i$ is the energy of the $i^{th}$ quasicontinuum state, and
summation is carried out symmetrically around $i=l$. The number of
averaging points $2N+1$ is optimized to reduce errors without
reducing significantly resolution on the energy scale for a smooth
distribution. Although we will illustrate the accuracy of the method
with quite specific atoms, the method can be used for obtaining
differential cross section distributions from quasi-continuum
discrete OSs more generally: for complex atoms and molecules and for
theories other than CI-RMBPT.

\begin{figure}
\centerline{\includegraphics*[scale=0.75]{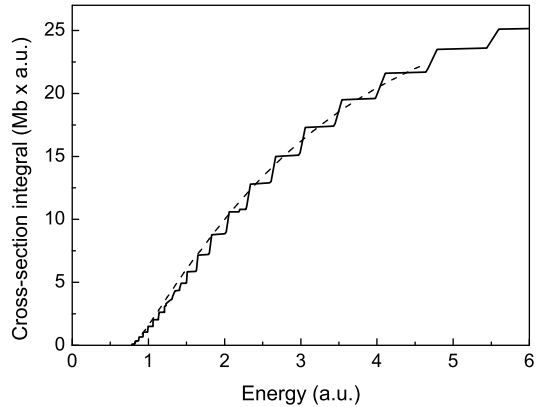}}
\caption{Comparison for neon of the integrated experimental
photoionization cross section of \cite{PrecNoble} (dashed line) with
the sum of quasicontinuum CI+RMBPT values (solid line)}
\label{IntNe}
\end{figure}

\begin{figure}
\centerline{\includegraphics*[scale=0.8]{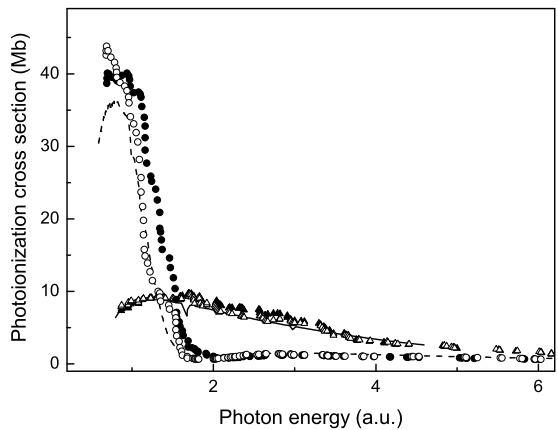}} \caption{Ne
and Ar photoionization cross sections. Ne data are under 10 Mb:
solid line - expt.\cite{PrecNoble}, open triangles - CI+RMBPT, solid
triangles - CI. Ar data: dashed line - expt.\cite{ChanEtAl3}, open
circles -  CI+RMBPT, solid circles - CI} \label{ArNe}
\end{figure}

In order to test the accuracy of the current photo-ionization theory
and conversion method, we calculated CI-RMPBT photo-ionization cross
sections for noble gas atoms from Ne to Xe and compared them with
experiment. In addition we present CI calculations to evaluate the
contribution from RMBPT part and estimate theoretical accuracy. The
oscillator strengths from the ground state to the quasicontinuum
states were converted to photo-ionization cross sections by using
line-fitting method and differentiation, as described in the
previous section. The number of data points for line fitting was
chosen to minimize ``noise" without substantial reduction in
resolution on the energy scale. This number depends on the distance
between quasicontinuum levels and hence the cavity size.

The results for neon are shown in Fig. 2 (the data below 10 Mb). The
PI cross section calculated with the CI-RMBPT method, expected to be
accurate in neon, agrees well with experiment for the whole range of
experimental energies. The CI results agree with experiment as well,
but CI-RMBPT values appear to be closer to the experimental cross
section. By extrapolating the difference between CI-RMBPT and CI
results, which is due to RMBPT correction, it is possible to
estimate theoretically that the CI-RMBPT calculation accuracy is on
the order of a few percent. Some ``fluctuations" of similar order in
theoretical calculations are present due to the conversion of
discrete OSs into cross sections. These fluctuations can be further
reduced by additional smoothing, but this reduces energy resolution.

\begin{figure}
\centerline{\includegraphics*[scale=0.75]{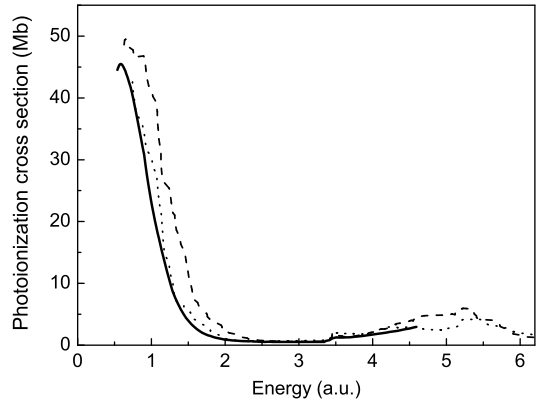}}
\caption{Krphotoionization cross sections: the solid line -
expt.\cite{PrecNoble}, the dotted line - CI+RMBPT, the dashed line -
CI calculations} \label{Kr}
\end{figure}

\begin{figure}
\centerline{\includegraphics*[scale=0.75]{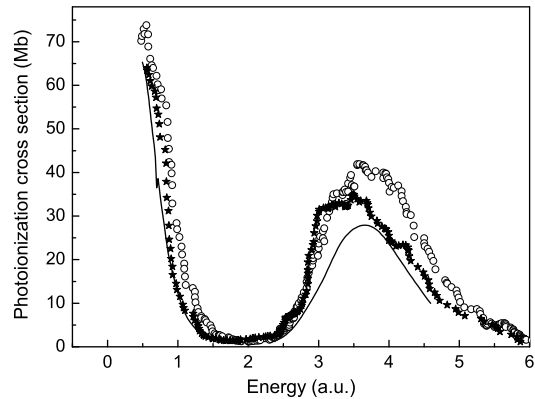}} \caption{Xe
photoionization cross sections: the solid line - Xe
expt.\cite{PrecNoble};  the stars - Xe CI+RMBPT, the open circles -
Xe CI.} \label{Xe}
\end{figure}
In Fig. 2, a comparison of CI+MBPT, CI, and experimental cross
sections is shown also for argon. The case of argon reveals much
better agreement of the experiment with the CI+MBPT than with the CI
theory, especially on the steep slope, with the agreement being
equally good at energies above 2 a.u. Substantial deviation of both
theories from experiment is observed at 1 a.u. energy, which can be
attributed to both inaccuracy of the conversion procedure and
neglected RMBPT terms. At the steep slope, the RMBPT contribution
leads to a shift of the curve to the left by about 0.2 a.u. The
shift due to neglected higher-order effects is expected to be
smaller, so we can conservatively state that the horizontal error is
about 0.1 a.u. The vertical error is about 5 Mb on the slope. The
conversion errors are smaller in this case than the error due to
neglect RMBPT terms.

In the case of Kr (Fig.3), the accuracy of CI+MBPT results is
similarly better than that of the CI results on the whole range of
shown energies. The RMBPT corrections in Kr resulting in the energy
shift of the PI curve have somewhat increased as expected for the
sequence and are clearly needed to be included. Finally the Xe cross
sections (Fig.4) also follow the trends of Ar and Kr that the slope
portion of the differential cross section is much better reproduced
with the CI-RMBPT than with the CI theory. However, the RMBPT shift
becomes smaller than in Ar and Kr. On the other hand, the Xe cross
section has a well pronounced second peak, which although is still
better reproduced with the CI-RMBPT than CI calculations, reveals
noticeable discrepancy between theory and experiment. This
discrepancy might be attributed to omitted higher-order RMBPT
effects.In general the theoretical accuracy can be roughly estimated
as the difference between CI and CI-RMPBT values, so the presented
calculations not only give more accurate values than the simpler CI
calculations but also theoretical errors.

 From the given demonstrations, we can summarize
that the CI+RMBPT theory gives higher accuracy than CI for noble-gas
atoms from Ne to Xe. This is quite an encouraging result, which not
only confirms that the CI+RMBPT method is an accurate theory of
atomic properties in the considered energy range, but also that the
conversion procedure described here can be used for accurate cross
section calculations. The developed method can be applied to many
complex atom where multiple irregular levels exist in the
quasi-continuum spectrum. In particular, the CI-RMBPT quasi-discrete
OSs can be converted  to cross sections in many atoms.

The work of I. Savukov has been performed with the support of Los
Alamos LDRD EC and under the auspices of the U.S.$\sim$DOE by LANL
under contract No.$\sim$DE-AC52-06NA25396. The work of D. Filin was
supported by LDRD EC through the New Mexico Consortium.

\end{document}